\documentclass[aip,reprint]{revtex4-1}

\draft 

\usepackage{amsmath,graphicx}
\usepackage{url}

\begin{document}


\title{Radio-frequency electromagnetic field and vortex penetration in multilayered superconductors} 

\author{Takayuki Kubo}
\email[]{kubotaka@post.kek.jp}
\affiliation{KEK, High Energy Accelerator Research Organization, 1-1 Oho, Tsukuba, Ibaraki 305-0801 Japan}

\author{Yoshihisa Iwashita}
\affiliation{Institute for Chemical Research, Kyoto University, Uji, Kyoto 611-0011 Japan}

\author{Takayuki Saeki}
\affiliation{KEK, High Energy Accelerator Research Organization, 1-1 Oho, Tsukuba, Ibaraki 305-0801 Japan}

\begin{abstract}
A multilayered structure with a single superconductor layer and a single insulator layer formed on a bulk superconductor is studied.
General formulae for the vortex-penetration field of the superconductor layer and the magnetic field on the bulk superconductor, 
which is shielded by the superconductor and insulator layers, are derived with a rigorous calculation of the magnetic field attenuation in the multilayered structure.
The achievable peak surface field depends on the thickness and its material of the superconductor layer, the thickness of the insulator layer and material of the bulk superconductor.
The calculation shows a good agreement with an experimental result. 
A combination of the thicknesses of superconductor and insulator layers 
to enhance the field limit can be given by the formulae for any given materials. 
\end{abstract}


\maketitle 

Technologies to fabricate the superconducting RF cavities made of Nb have been advanced.  
The maximum accelerating gradient $E_{\rm acc}$ of the TESLA type $1.3\,{\rm GHz}$ 9-cell cavities during performance tests in vertical cryostats regularly exceed $35\,{\rm MV/m}$ at several laboratories.  
The gradient record had been increasing and recently two 9-cell cavities made from large grain Nb reached $45\,{\rm MV/m}$ at DESY~\cite{singer}. 
Further high gradients, however, would not be expected because their gradients are thought to be close to the empirical limit imposed by the thermodynamic critical field $\simeq 200\, {\rm mT}$ of Nb~\cite{hassan}. 
A. Gurevich suggested~\cite{gurevich, gurevichreview} that a multilayered nanoscale coating on Nb cavity may push up the RF breakdown field to the level of the vortex-penetration field of the coating materials 
at which the Bean-Livingston surface barrier~\cite{beanlivingston} disappears.  
While some experimental studies have been conducted on the subject based on the idea~\cite{tajima, antoine}, not much theoretical progress followed on it.
In fact, the best parameter set for the multilayer coating model such as thicknesses of layers and choices of materials are not clear from a theoretical point of view. 
In this letter, the multilayered structure is carefully evaluated with a rigorous calculation on the electromagnetic field distribution to keep its self-consistency. 
The resultant vortex-penetration field, the best combination of parameters, and materials are described.

\begin{figure}[tb]
   \begin{center}
   \includegraphics[width=0.9\linewidth]{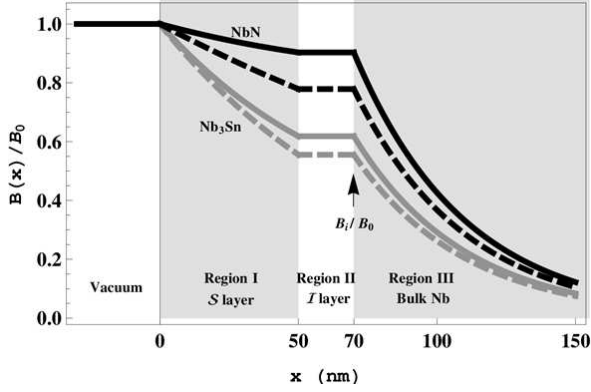}
   \end{center}\vspace{-0.6cm}
   \caption{
Examples of the magnetic field attenuations in the multilayered structure.  
Solid curves show our formulae given above, and dashed curves show the naive estimates with $B=B_0 e^{-x/\lambda_1}$.
Black curves and gray curves correspond to the material of the ${\mathcal{S}}$ layer: 
${\rm NbN}$ ($\lambda_1=200\,{\rm nm}$) and ${\rm Nb}_3{\rm Sn}$ ($\lambda_1=85\,{\rm nm}$), respectively. 
The bulk superconductor is assumed to be Nb ($\lambda_2=40\,{\rm nm}$). 
The values of $\lambda_1$ and $\lambda_2$ are given in literature~\cite{gurevichreview}. 
The thickness of the ${\mathcal{S}}$ layer and the ${\mathcal{I}}$ layer are fixed at $d_{\mathcal{S}}=50\,{\rm nm}$ and $d_{\mathcal{I}}=20\,{\rm nm}$. 
   }\label{figure1}
\end{figure}

The multilayer coating model~\cite{gurevich} consists of alternating layers of superconductor layers ($\mathcal{S}$) and insulator layers ($\mathcal{I}$). 
The simplest configuration with a single superconductor layer and a single insulator layer is seen in Fig.~\ref{figure1}.  
Each $\mathcal{S}$ layer is expected to withstand higher field than bulk Nb, 
and to shield the bulk Nb from the applied RF surface magnetic field $B_0$, 
because $B_i$ (the RF surface field on the bulk Nb) is smaller than $B_0$. 
Then the multilayered structure is thought to withstand a higher field than the bulk Nb   
if $B_0$ is smaller than the vortex-penetration fields of the top $\mathcal{S}$ layer and $B_i$ is smaller than that of the bulk Nb.  
The vortex-penetration field of the ${\mathcal{S}}$ layer was given by $B_v = \phi_0/4\pi \lambda \xi$ in the original paper~\cite{gurevich}, 
where $\phi_0=2.07\times 10^{-15}\,{\rm Wb}$ is the flux quantum~\cite{tinkham}, and $\lambda$ and $\xi$ are a London penetration depth and a coherence length of the material of the ${\mathcal{S}}$ layer, respectively.  
This expression, however, has the same form as the vortex-penetration field of the semi-infinite superconductor, 
and does not depend on any parameters on the configuration of the multilayered structure such as the ${\mathcal{S}}$ layer thickness or the ${\mathcal{I}}$ layer thickness. 
In order to incorporate effects from the configuration of multilayered structure, 
we carried out rigorous caluculation on the distribution of magnetic field and Meissner current in the ${\mathcal{S}}$ layer.

\begin{table*}[tb]
\caption{\label{tab:table1}
Parameters of the multilayered structure with a single superconductor layer and a single insulator layer formed on a bulk superconductor.
}
\begin{ruledtabular}\label{table1}
\begin{tabular}{lcc}
Region & Material type   & Parameter \\
\colrule
I   & Superconductor layer & Coherence length: $\xi_1$, London penetration depth: $\lambda_1$ ($\gg \xi_1$), Thickness: $d_{\mathcal{S}}$ ($\gg \xi_1$) \\
II  & Insulator layer      & Relative permittivity: $\epsilon_r$, Thickness: $d_{\mathcal{I}}$ (zero or larger than a few nm) \\
III & Bulk superconductor  & London penetration depth: $\lambda_2$ \\
\end{tabular}
\end{ruledtabular}
\end{table*}

In order to derive the electromagnetic field in the multilayered structure, 
the Maxwell equations and the London equations should be solved with appropriate boundary conditions simultaneously. 
Contributions to the electromagnetic field distribution from the normal (unpaired) electrons of the superconductor and dielectric losses in the insulator are neglected. 
For simplicity, let us consider a model with a single $\mathcal{S}$ layer and a single $\mathcal{I}$ layer formed on a bulk superconductor as shown in Fig.~\ref{figure1}. 
Table.~\ref{table1} shows the parameters for the model. 
$d_{\mathcal{I}}$ is assumed to be zero or larger than a few ${\rm nm}$ to suppress the Josephson coupling~\cite{gurevichreview}. 
All layers are parallel to the $y$-$z$ plane and then perpendicular to the $x$-axis. 
The electric and magnetic fields are assumed to be parallel to the layers: 
${\bf E}= (0,E,0) e^{-i\omega t}$ and ${\bf B}= (0,0,B) e^{-i\omega t}$, 
where $E$ and $B$ are amplitudes of electric and magnetic fields, and $\omega$ is an angular frequency. 
Further we assume 
the materials used for the ${\mathcal{S}}$ layer is extreme Type II superconductor $\lambda_1 \gg \xi_1$, and 
the ${\mathcal{S}}$ layer thickness is larger than the coherence length $d_{\mathcal{S}} \gg \xi_1$. 
Note that the ${\mathcal{S}}$ layer of our model is not necessarily a thin film 
hence the discussion below can be applied to any ${\mathcal{S}}$ layer with arbitrary thickness $d_{\mathcal{S}} \gg \xi_1$. 
Solving the Maxwell equations in the ${\mathcal{I}}$ layers, and the London equations in the ${\mathcal{S}}$ layers and in the bulk superconductor 
with continuity conditions of electric and magnetic fields at boundaries~\cite{ipac13kis}, 
we find 
\begin{eqnarray}
B_{\rm I} &=& 
B_0
\frac{ \cosh \frac{d_{\mathcal{S}}-x}{\lambda_1} + (\frac{\lambda_2}{\lambda_1} + \frac{d_{\mathcal{I}}}{\lambda_1}) \sinh \frac{d_{\mathcal{S}}-x}{\lambda_1}}
     { \cosh \frac{d_{\mathcal{S}}}{\lambda_1} + (\frac{\lambda_2}{\lambda_1} + \frac{d_{\mathcal{I}}}{\lambda_1}) \sinh \frac{d_{\mathcal{S}}}{\lambda_1}} \,  ,  
\label{eq1} 
\end{eqnarray}
\begin{eqnarray}
B_{\rm II} &=& 
B_0
\frac{ 1 }
     { \cosh \frac{d_{\mathcal{S}}}{\lambda_1} + (\frac{\lambda_2}{\lambda_1} + \frac{d_{\mathcal{I}}}{\lambda_1}) \sinh \frac{d_{\mathcal{S}}}{\lambda_1}} \,  ,  
\label{eq2} 
\end{eqnarray}
\begin{eqnarray}
B_{\rm III} &=& 
B_0 \frac{ e^{-\frac{x-d_{\mathcal{S}} -d_{\mathcal{I}}}{\lambda_2}}}
         { \cosh \frac{d_{\mathcal{S}}}{\lambda_1} + (\frac{\lambda_2}{\lambda_1} + \frac{d_{\mathcal{I}}}{\lambda_1}) \sinh \frac{d_{\mathcal{S}}}{\lambda_1}} \, , 
\label{eq3}
\end{eqnarray}
where $B_{\rm I}$, $B_{\rm II}$, and $B_{\rm III}$ are the amplitudes of magnetic fields in region I, II, and III, respectively.
Note here that these equations are approximated formulae that are valid for $d_{\mathcal{I}} \ll (\sqrt{\epsilon_r}k)^{-1}$,  
where $k=\omega/c$, and $c$ is the speed of light. 
For example, 
a frequency $f=\omega/2\pi=1.3\,{\rm GHz}$~\cite{TDR} imposes $d_{\mathcal{I}} \ll 1\,{\rm cm}$ as a condition of validity. 
It is easy to confirm that these equations are reduced to the well known expression for the semi-infinite superconductor given by $B=B_0 e^{-x/\lambda_1}$  
when the ${\mathcal{S}}$ layer and the bulk superconductor are the same material ($\lambda_1 = \lambda_2$) and the ${\mathcal{I}}$ layer vanishes ($d_{\mathcal{I}} \to 0$). 
Fig.~\ref{figure1} shows examples how a magnetic field attenuates in a multilayered structure.

\begin{figure}[tb]
   \begin{center}
   \includegraphics[width=0.9\linewidth]{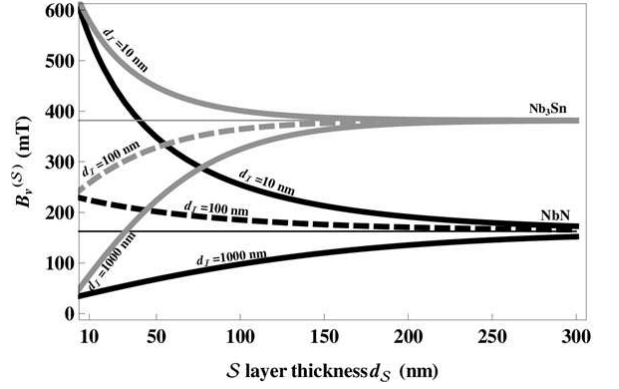}
   \end{center}\vspace{-0.6cm}
   \caption{
Vortex-penetration fields of the ${\mathcal{S}}$ layer, $B_v^{(\mathcal{S})}$, as functions of an ${\mathcal{S}}$ layer thickness $d_{\mathcal{S}}$.  
Solid curves, dashed curves and dashed-dotted curves correspond to ${\mathcal{I}}$ layer thickness 
$d_{\mathcal{I}}=10\,{\rm nm}$, $100\,{\rm nm}$, and $1000\,{\rm nm}$, respectively. 
Thin horizontal lines represents asymptotic lines, which correspond to vortex-penetration fields of thick ${\mathcal{S}}$ layers ($d_{\mathcal{S}} \gg \lambda_1$). 
Black curves and gray curves correspond to the material of the ${\mathcal{S}}$ layer: 
${\rm NbN}$ ($\lambda_1=200\,{\rm nm}$, $\xi_1=5\,{\rm nm}$) and 
${\rm Nb}_3{\rm Sn}$ ($\lambda_1=85\,{\rm nm}$, $\xi_1=5\,{\rm nm}$), respectively. 
The bulk superconductor is assumed to be Nb ($\lambda_2=40\,{\rm nm}$). 
The values of $\xi_1$ are derived from $\phi_0/(2\sqrt{2}\pi \lambda_1 \xi_1)=B_c$, where the critical field $B_c$ of each material is given in literature~\cite{gurevichreview}. 
   }\label{figure2}
\end{figure}

The vortex-penetration field can be evaluated by computing two forces acting on a vortex at a top of the ${\mathcal{S}}$ layer: 
a force from an image current of an image antivortex which is introduced to satisfy a boundary condition of zero current normal to the surface, and 
another from a Meissner current ${\bf j}_{\rm M}$ due to existence of external field which can be computed from Eq.~(\ref{eq1}) with $j_{{\rm M}y}=-(1/\mu_0)dB_{\rm I}/dx$.  
Then the vortex-penetration field is given by~\cite{srf13kis}
\begin{eqnarray}
B_v^{(\mathcal{S})}  
&=& \frac{\phi_0}{4\pi \lambda_1 \xi_1} 
  \frac{\cosh\frac{d_{\mathcal{S}}}{\lambda_1} + (\frac{\lambda_2}{\lambda_1} + \frac{d_{\mathcal{I}}}{\lambda_1})\sinh\frac{d_{\mathcal{S}}}{\lambda_1}}
       {\sinh\frac{d_{\mathcal{S}}}{\lambda_1} + (\frac{\lambda_2}{\lambda_1} + \frac{d_{\mathcal{I}}}{\lambda_1})\cosh\frac{d_{\mathcal{S}}}{\lambda_1} } \,, \label{eq4} 
\end{eqnarray}
which depends on both the ${\mathcal{S}}$ layer thickness $d_{\mathcal{S}}$ and the ${\mathcal{I}}$ layer thickness $d_{\mathcal{I}}$. 
Note here that Eq.~(\ref{eq4}) is reduced to the well-known expression $\phi_0/4\pi \lambda_1 \xi_1\, (\equiv B_v^{(\mathcal{S}_{\infty})})$ for the semi-infinite ${\mathcal{S}}$ layer ($d_{\mathcal{S}}\to \infty$).  
As is obvious from Eq.~(\ref{eq4}) and Fig.~\ref{figure2}, $B_v^{(\mathcal{S})}$ increases to $(\lambda_1/\lambda_2)B_v^{(\mathcal{S}_{\infty})}$ as $d_{\mathcal{S}}$ and $d_{\mathcal{I}}$ decrease. 
This behavior can be understood from the above results that 
the magnetic field less attenuates in a thin ${\mathcal{S}}$ layer on a thin ${\mathcal{I}}$ layer.  
This means that a Meissner current, which is proportional to a gradient of the magentic field, becomes smaller as $d_{\mathcal{S}}$ and $d_{\mathcal{I}}$ decrease, 
and a force that draw a vortex into the ${\mathcal{S}}$ layer becomes weak. 
As a result, a field that the ${\mathcal{S}}$ layer can withstand, $B_v^{(\mathcal{S})}$, increases.

\begin{table}[tb]
\caption{\label{tab:table2}
Summary of optimum parameters, $d_{\mathcal S}$ and $d_{\mathcal I}$, and resultant $B_v^{\rm (ML)}$. 
}
\begin{tabular}{|c|c|c|}
\hline
${\mathcal S}$ layer            & \multicolumn{2}{c|}{bulk superconductor ($\lambda_2\,, B_v^{\rm (bulk)}$)} \\ \cline{2-3}
 ($\lambda_1,\,\xi_1$)          & Nb ($40 \,{\rm nm}\,, 200\,{\rm mT}$)  & ${\rm Nb}^*$ ($300 \,{\rm nm}\,, 20\,{\rm mT}$)  \\
\hline
${\rm NbN}$                     & $B_v^{\rm (ML)}=240\,{\rm mT}$         & $B_v^{\rm (ML)}=160\,{\rm mT}$  \\
$(200\,{\rm nm},\,5\,{\rm nm})$ & $d_{\mathcal{S}}=100\,{\rm nm}$        & $d_{\mathcal{S}}\gg \lambda_1$ \\
                                & $d_{\mathcal{I}}\lesssim 20\,{\rm nm}$ & $d_{\mathcal{I}}=$ arbitrary \\
\hline
${\rm MgB_2}$                   & $B_v^{\rm (ML)}=300\,{\rm mT}$         & $B_v^{\rm (ML)}=230\,{\rm mT}$ \\
$(140\,{\rm nm},\,5\,{\rm nm})$ & $d_{\mathcal{S}}=70\,{\rm nm}$         & $d_{\mathcal{S}}\gg \lambda_1$  \\
                                & $d_{\mathcal{I}}\lesssim 20\,{\rm nm}$ & $d_{\mathcal{I}}=$ arbitrary \\
\hline
${\rm Nb_3Sn}$                  & $B_v^{\rm (ML)}=400\,{\rm mT}$         & $B_v^{\rm (ML)}=380\,{\rm mT}$ \\
$(85\,{\rm nm},\,5\,{\rm nm})$  & $d_{\mathcal{S}}=90\,{\rm nm}$         & $d_{\mathcal{S}}\gg \lambda_1$  \\
                                & $d_{\mathcal{I}}\lesssim 20\,{\rm nm}$ & $d_{\mathcal{I}}=$ arbitrary \\
\hline
\end{tabular}
\end{table}

\begin{figure}[tb]
   \begin{center} 
   \includegraphics[width=0.9\linewidth]{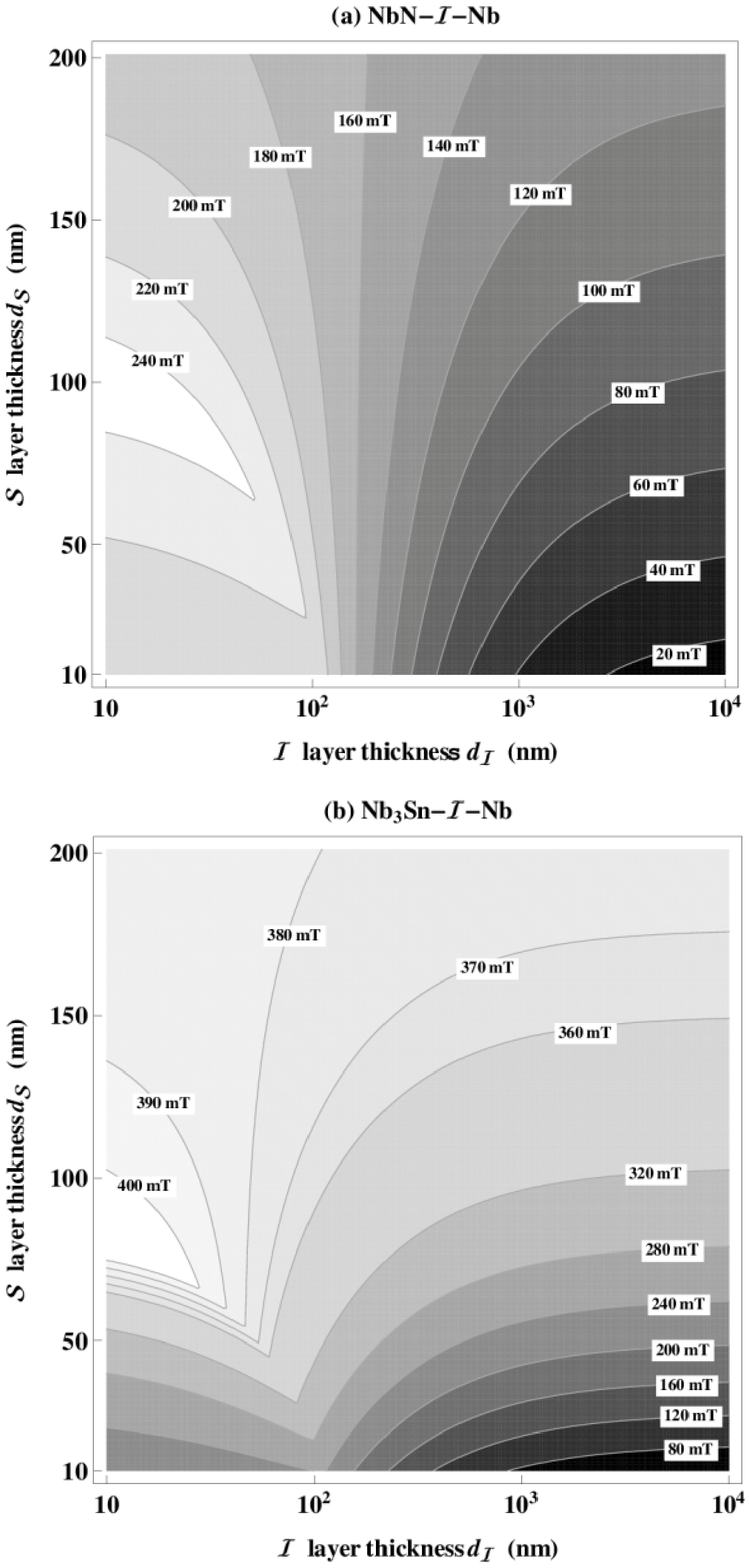} 
   \end{center}\vspace{-0.6cm}
   \caption{
Contour plots of the maximum achievable peak surface-field without vortex dissipations, $B_v^{\rm (ML)}$. 
The abscissa represents the ${\mathcal{I}}$ layer thickness $d_{\mathcal{I}}$ and 
the ordinate represents the ${\mathcal{S}}$ layer thickness $d_{\mathcal{S}}$. 
Values written in the plot area are $B_v^{\rm (ML)}$ in the unit of mT.  
The top and bottom figures correspond to materials of the ${\mathcal{S}}$ layer. 
(a) ${\rm NbN}$ ($\lambda_1=200\,{\rm nm}$, $\xi_1=5\,{\rm nm}$) and 
(b) ${\rm Nb}_3{\rm Sn}$ ($\lambda_1=85\,{\rm nm}$, $\xi_1=5\,{\rm nm}$), respectively. 
The bulk superconductor is assumed to be Nb with $\lambda_2=40\,{\rm nm}$ and $B_v^{\rm bulk}=200\,{\rm mT}$. 
   }\label{figure3}
\end{figure}

\begin{figure}[tb]
   \begin{center} 
   \includegraphics[width=0.9\linewidth]{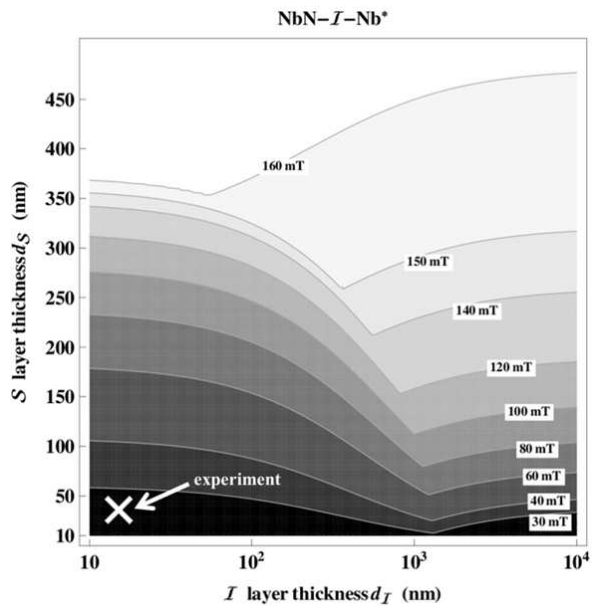} 
   \end{center}\vspace{-0.6cm}
   \caption{
A Contour plot of the maximum achievable peak surface-field without vortex dissipations, $B_v^{\rm (ML)}$. 
The abscissa represents the ${\mathcal{I}}$ layer thickness $d_{\mathcal{I}}$ and 
the ordinate represents the ${\mathcal{S}}$ layer thickness $d_{\mathcal{S}}$. 
Values written in the plot area are $B_v^{\rm (ML)}$ in the unit of mT.  
Materials of the ${\mathcal{S}}$ layer and the bulk superconductor are assumed to be 
${\rm NbN}$ ($\lambda_1=200\,{\rm nm}$, $\xi_1=5\,{\rm nm}$) and 
magnetron sputtered Nb ($\lambda_2=300\,{\rm nm}$~\cite{balalykin} and $B_v^{\rm bulk}=20\,{\rm mT}$~\cite{antoine2}), respectively. 
A cross shown at the lower left indicate a parameter set used in an experiment~\cite{antoine2}. 
   }\label{figure4}
\end{figure}

A thin ${\mathcal{S}}$ layer pushes up $B_v^{(\mathcal{S})}$, but it can not protect the bulk superconductor from an applied field if $d_{\mathcal{S}} \ll \lambda_1$. 
In order to evaluate the achievable peak surface-field without vortex dissipations, 
not only $B_v^{(\mathcal{S})}$, but also the shielded magnetic field on the bulk superconductor must be considered simultaneously. 
Let us define the magnetic field attenuation ratio $\alpha$ by $\alpha=B_{\rm II}/B_0$.    
When the magnetic field attenuation in the ${\mathcal{S}}$ layer is enough for the shielded magnetic field $\alpha B_v^{(\mathcal{S})}$ to become smaller than a vortex penetration field of the bulk superconductor, $B_v^{\rm (bulk)}$, the bulk superconductor is safely protected. 
Then the achievable peak surface-field without vortex dissipations, $B_v^{\rm (ML)}$, is given by $B_v^{(\mathcal{S})}$. 
On the other hand when the magnetic field attenuation is not enough and $\alpha B_v^{(\mathcal{S})}$ is larger than $B_v^{\rm (bulk)}$, 
$B_v^{\rm (ML)}$ is limited by $\alpha^{-1} \times B_v^{\rm (bulk)}$. 
Thus we find 
\begin{eqnarray}\label{eq5}
B_v^{\rm (ML)} =
  \begin{cases}
    B_v^{(\mathcal{S})}                & (\alpha B_v^{(\mathcal{S})} < B_v^{\rm (bulk)}) \\
    \alpha^{-1} \times B_v^{\rm (bulk)}  & (\alpha B_v^{(\mathcal{S})} \ge B_v^{\rm (bulk)}) . 
  \end{cases}
\end{eqnarray}
Fig.~\ref{figure3} shows $B_v^{\rm (ML)}$ of (a) NbN-${\mathcal I}$-Nb structure and (b) ${\rm Nb_3Sn}$-${\mathcal I}$-${\rm Nb}$ structure. 
A choice of  appropriate parameter regions should improve $B_v^{\rm (ML)}$: 
a combination of an NbN layer (${\rm Nb_3Sn}$ layer) with $d_{\mathcal{S}}=100\,{\rm nm}$ ($90\,{\rm nm}$) and 
an ${\mathcal{I}}$ layer with $d_{\mathcal{I}} = 10\,{\rm nm}$
yields $B_v^{\rm (ML)} \simeq 240\,{\rm mT}$ ($400\,{\rm mT}$). 
$B_v^{\rm (ML)}$ of NbN-${\mathcal I}$-${\rm Nb}^*$ structure is shown in Fig.~\ref{figure4}, 
where ${\rm Nb}^*$ represents a magnetron sputtered Nb. 
A thick ${\mathcal{S}}$ layer ($d_{\mathcal{S}}\gg \lambda_1$) with arbitrary $d_{\mathcal{I}}$ yields the maximum $B_v^{\rm (ML)}=160\,{\rm mT}\, (= B_v^{\mathcal (S_{\infty})})$. 
A thin ${\mathcal{S}}$ layer yields a rather small $B_v^{\rm (ML)}$. 
In general, a bulk superconductor with $\lambda_2>\lambda_1$, such as ${\rm Nb}^*$, suppresses $B_v^{(\mathcal{S})}$ (see Eq.~(\ref{eq4})) and thus $B_v^{\rm (ML)}$. 
Table.~\ref{tab:table2} summarizes optimum parameters and resultant maximum $B_v^{\rm (ML)}$.

On measurements of $B_v^{\rm (ML)}$, the magnetic field must be applied on one side of the layers. 
An experiment~\cite{antoine2} shows $B_v^{\rm (ML)} \simeq 30\,{\rm mT}$ for the case of 
NbN($25\,{\rm nm}$)-MgO($14\,{\rm nm}$)-${\rm Nb}^*$($250\,{\rm nm}$), 
which agrees well with the above calculation (see Fig.~\ref{figure4}). 
Increasing $d_{\mathcal{S}}$ or using regular Nb instead of ${\rm Nb}^*$ might drastically improve $B_v^{\rm (ML)}$.

These calculations are performed on the ideal superconductor and insulator. 
In real situations, however, the superconductor includes defects and surface roughnesses, 
and both layers have fluctuations in thickness. 
Effects of these complicated conditions should be considered in the next step.

As for geometrical conditions, only the electromagnetic field propagates perpendicular to the surface of the layers are considered in this article.
When the normal components have non-zero value, 
additional resonance modes associated with the standing waves confined in the insulator layer would appear. 
Since the extent of the insulator layer could be as long as the wave length of the operating frequency of the cavity, 
it is possible that additional resonance modes emerge near the operating frequency. 
In addition to the above points, variations of geometry and electromagnetic fields in other directions should be considered in accordance with real accelerating cavities. 
Study on effects from these additional conditions, however, is a future challenge.


\newpage


\end{document}